\documentstyle{article}
\title{Entangled Simultaneous Measurement and
Elementary Particle Representations}
\author{Robert Y. Levine$^{*}$ \and Victor Dannon$^{**}$ \\
Spectral Sciences, Inc., 99 S. Bedford St.,
Burlington, MA. 01803$^{*}$
\\California State University,
 San Marcos CA 92096$^{**}$ }
\date{May 17, 2000}
\begin{document}
\maketitle

\large

\begin{abstract}
\normalsize
It is proposed that the principles of relativistic quantum mechanics are
incomplete for simultaneous measurement of non-commuting operators.
Consistent joint
measurement of incompatible observables at a single point in space-time
requires that the system be in an entangled state with vacuum meters.
Entangled simultaneous measurement for non-commuting observables is
suggested as the basis for observed fermionic multiplets.
This generalizes the standard spin representations for particles
arising from Lorentz invariance. It is shown
that operator entanglement for all quantum observables in the Poincare
algebra, coupled with Fermi-Dirac statistics, mandates six fermions.
The quark and lepton generations are proposed to form a super-structure
of the Poincare algebra based on the principle of entangled simultaneity.
Mathematically, the structure is known as a Naimark extension.
The required entanglement between particle generations for left-handed quarks
is observed in the Cabibbo-Kobayashi-Maskawa matrix. In the
appendix, the Naimark-extended von Neumann lattice is shown to be
distributive, thereby suggesting the principle of entangled simultaneity as
a mechanism to avoid quantum non-locality.
 Levine, bob@spectral.com.
\end{abstract}

\section{Introduction}

The measurement of non-commuting operators in a quantum
system is at the root of the well-known paradoxes of quantum mechanics. For
example, the Kochen-Specker paradox involves measurement of a set of non-commuting
operators with mutually commuting subsets \cite{kochenspecker,mermin}. The
Bell inequalities are similarly formulated by incompatible space-like separated
spin measurements \cite{mermin,bell}. The connection of these problems
to quantum non-locality
motivates the consideration of an alternative
definition for the measurement of incompatible quantum observables.
Because measured quantities
are inherently statistical, quantum mechanics requires an ensemble of identical
systems to establish an expectation value. The types of measurements on the
ensemble in the situation of non-commuting observables is critical to the
interpretation of the result. For example, the usual simultaneous measurement
of position and momentum would require separate measurements of each operator
on half the ensemble at the same time. This procedure for the assignment of
position and momentum to a system, coupled with wavefunction collapse, leads
to a built-in non-locality in system observables. One approach to this problem
is to restrict {\em valid} measurement of non-commuting observables to be with
special-purpose ancillary quantum systems (referred to as vacuum meters in this
paper) that are entangled with the original system. In the alternative {\em
entangled simultaneous} measurement scheme, all systems in the ensemble have
the same experimental set-up with coupling to vacuum meters for a joint
position/momentum measurement. The premise of this paper is that this procedure,
while avoiding quantum non-locality, yields results that are properly
interpreted as a joint measurement of non-commuting observables. Furthermore,
it is assumed that, probably due to a principle involving quantum non-locality,
entangled states are fundamental to particle representations. The principle
mandates that particle states allow the simultaneous determination of all
particle observables. These assumptions impose the structure of a Naimark extension
on particle multiplets.

Allowed quantum numbers and statistics for relativistic particles are strictly
constrained by the dual principles of Lorentz invariance and quantum mechanics
\cite{weinberg}. The latter requires states that represent symmetries of the
lagrangian. A larger structure, the Naimark extension of the Poincare algebra
\cite{naimarka,naimarkb}, results if states are required for the measurement
of incompatible observables with entangled vacuum meters. New commuting operators
are defined in a Naimark extension that project to the original set upon meter
measurement. A key component of the measurement scheme is the entanglement of the
system with a vacuum state containing independent meters. The assumption of
Fermi-Dirac statistics for the meters forces a different particle identity (flavor)
onto the ancillary particles that make up the meters. It is shown in this paper that
the minimum Hilbert space for the Naimark extension of the Poincare algebra
contains six independent fermions, which are identified with left- and right-handed
lepton/quark generations. The particle set is the minimum required for
realization of the Poincare algebra on commuting operators such that particles can
entangle with vacuum meters. The universal nature of fermionic multiplets,
existing for both leptons and quarks, motivates the suggestion of a single
underlying principle rooted in relativistic quantum mechanics. An elementary
particle is interpreted as a complex, entangled quantum system in which the
entire space of Poincare observables is realized on commuting operators.

The Naimark extension or embedding is a mathematical description of measurement with
a quantum apparatus \cite{vonNeumann}. In the original construction non-orthogonal
projection operators, such as generated by optical coherent states \cite{helstrom},
are extended to orthogonal projection operators in a combined system/meter
Hilbert space \cite{davies}. As first proposed by von Neumann \cite{vonNeumann},
and developed by Arthurs and Kelly \cite{arthurskelly}, a realization of the
Naimark extension for position and momentum is obtained by the entangling of a
harmonic oscillator with meter harmonic oscillators in ground states. In the
Arthurs-Kelly model, measurement on two meters results in the collapse of the
system wavefunction to a coherent state corresponding to the measured position
and momentum. As explained by Levine and Tucci \cite{levinetuccia}, entangled
simultaneous measurement of position and momentum with a single meter results
in the collapse of the system to the eigenstate of the operator that was
measured on the system. However, the system/meter expectation values are still
proportional to the desired system expectation values. The theory of entangled
simultaneous quantum measurement was extended to non-relativistic spin by
coupling to spin-1/2 meters by Levine and Tucci \cite{levinetuccib}. In this case
measurements project the system to Bloch states corresponding to the measured
spin components. Analogous simultaneous spin measurement schemes are found in
Refs. \cite{buscha}-\cite{buschschroeck}. Modified Stern-Gerlach
experiments, based on hamiltonian models for simultaneous spin measurement,
are discussed in Ref. \cite{martins}. The relativistic generalization of
position, momentum, and angular momentum measurements leads to the consideration
of the Poincare algebra of observables and elementary particles. In an attempt
to describe elementary particle multiplets, Levine \cite{levinea} suggested the
generalization of spin $(SU(2))$ measurement to the measurement of $SO(2n)$
Clifford algebra operators using $2n$ spin meters. A more fundamental application
to elementary particle multiplets, originating in the structure of the Poincare
algebra, is proposed in this paper. Finally, a general discussion of entangled
simultaneous measurement with second quantized relativistic fields is found
in Ref. \cite{levineb}.

The mechanism for entangled simultaneous measurement is particularly transparent
for second quantized systems. The vacuum, which is defined as the state projected
to zero by annihilation operators, is critical to isolate incompatible
system observables on independent meters.
Another key property of the Naimark extension is that
the minimal extension is determined from the pattern of commutativity in the
algebra of observables. These properties are demonstrated for non-relativistic
position, momentum, and angular momentum measurements in Section~2. In the
latter case the three components of angular momentum are measured through
the entanglement of the system with two vacuum meters. The formalism for first
quantized harmonic oscillators, reviewed in the section, is directly generalized
to relativistic fields. Section~3 contains the relativistic generalization of
simultaneous measurement for Dirac fermions. It is shown that the Naimark extension
of the Poincare algebra consists of six independent fermionic fields. The
structure results from the embedding of the three 3-vector operators of momentum,
angular momentum, and boost generators. It is suggested that the three
entangled lepton/quark generations, an entanglement that is described for
left-handed quarks by the Cabibbo-Kobayashi-Maskawa matrix
\cite{kobayashimaskawa}, is a result of the Naimark extension of the Poincare
algebra for massive fermions. Appendices~A and B contain
the formal details of Naimark extensions for second quantized angular momentum
and Dirac fields, respectively.

As mentioned above, non-commuting operators are the basis for observed
non-locality in quantum systems. The quantum logical consequence of
incompatible operators is the non-distributivity of the von Neumann
subspace lattice of quantum measurement outcomes
\cite{vonNeumann,hughesa}. The possibility of a deeper reality,
involving a new principle of quantum measurement, motivates the suggestion that
measurement
of non-commuting observables must be confined to Naimark-extended Hilbert spaces. In
Appendix~C, it is shown that this principle of entangled simultaneity can be
applied to the subspace lattice to avoid paradoxical (non-distributive)
statements. As an example, the distributive Naimark-extended
lattice is constructed for spin-1/2 measurement. A conclusion follows in Section~4.

\section{Non-relativistic Systems}

In this section the simultaneous measurement of position and momentum in a first
quantized harmonic oscillator, and of angular momentum in a second
quantized system, are considered. Both cases involve the construction of
non-relativistic Naimark extensions, and have properties that generalize to the
relativistic case.

An example of a non-relativistic Naimark extension without second quantization
uses a pair of one-dimensional harmonic oscillators \cite{levinetuccib} with
the hamiltonian $(\hbar=c=1)$
\begin{equation}
H= \sum_{j=1}^{2} (\frac{p_{j}^{2}}{2m_{j}} + \frac{1}{2} m_{j} \omega_{j}^{2}
                                                          q_{j}^{2}).
\label{hamiltonian}
\end{equation}
In terms of annihilation operators,
$a_{j}=\sqrt{m_{j} \omega_{j}/2} (q_{j} + i p_{j}/m_{j} \omega_{j})$, the expression
in Eq.~(\ref{hamiltonian}) is written,
\begin{equation}
H=\sum_{j} \omega_{j} a_{j}^{\dag} a_{j} = \vec{a}^{\dag} D \vec{a},
\label{hamrepresent}
\end{equation}
where $\vec{a}^{\dag}=(a_{1}^{\dag},a_{2}^{\dag})$,$D=diag(\omega_{1},\omega_{2})$,
and the constant zero point energy is dropped. A rotation by angle $\theta$,
$\vec{A}=R(\theta)\cdot\vec{a}$, is applied to these operators to define another
Hilbert space with combined hamiltonian, $H=H_{0}+H_{int}$, given by
\begin{equation}
H_{0}=(c^{2}\omega_{1} + s^{2}\omega_{2})A_{1}^{\dag}A_{1} +
      (s^{2}\omega_{1} + c^{2}\omega_{2})A_{2}^{\dag}A_{2},
\label{ham0}
\end{equation}
and
\begin{equation}
H_{int}=sc(\omega_{1}-\omega_{2})(A_{1}^{\dag}A_{2} + A_{2}^{\dag}A_{1}),
\label{hamint}
\end{equation}
with $s=\sin\theta$ and $c=\cos\theta$. Note that, from the commutativity of
$a_{1}$ and $a_{2}$, the operators $A_{1}$ and $A_{2}$ commute. Through this
simple rotation, the operators $a_{j}(t)=e^{-i \omega_{j} t}a_{j}(0)$,
$j=1,2$, combine incompatible information about the $A_{1}$ system into
operators that are simultaneously measured. For example, consider the
commuting operators $q_{1}(t)=(a_{1}(t)+a_{1}^{\dag}(t))/\sqrt{2m_{1}\omega_{1}}$,
and $p_{2}(t)=-i\sqrt{m_{2}\omega_{2} /2}(a_{2}(t)-a_{2}^{\dag}(t))$ for system 1
position and system 2 momentum, respectively. In terms of the operators
$A_{1}(0)$ and $A_{2}(0)$ at time zero, these operators are given by
\begin{equation}
q_{1}(t)=\frac{1}{\sqrt{2m_{1}\omega_{1}}}(e^{-i\omega_{1} t}(c A_{1}(0) + s A_{2}(0))
                   +     e^{i\omega_{1} t} (c A_{1}(0) + s A_{2}(0))^{\dag}),
\label{position1}
\end{equation}
and
\begin{equation}
p_{2}(t)=i \sqrt{m_{2}\omega_{2} /2}(e^{-i\omega_{2}t}(s A_{1}(0) - cA_{2}(0))
               -  e^{i \omega_{2} t}(s A_{1}(0) - c A_{2} (0))^{\dag}).
\label{momentum2}
\end{equation}
The expectation of the operators $q_{1}(t)$ and $p_{2}(t)$ in the state
$|\psi>_{A_{1}} |0>_{A_{2}}$, in which the $A_{2}$ system is in the vacuum, is
given by
\begin{eqnarray}
<q_{1}(t)> & = & \frac{c}{\sqrt{2 m_{1} \omega_{1}}}
[\cos(\omega_{1} t)<\psi|(A_{1}(0)+A_{1}^{\dag}(0))|\psi> - \nonumber \\
& & i\sin(\omega_{1} t)<\psi|(A_{1}(0) - A_{1}^{\dag}(0))|\psi>]
\label{posexp}
\end{eqnarray}
and
\begin{eqnarray}
<p_{2}(t)>& = & -is \sqrt{\frac{m_{2} \omega_{2}}{2}}
[i \sin(\omega_{2} t)<\psi|(A_{1}(0)+A_{1}^{\dag}(0))|\psi> - \nonumber \\
& & \cos(\omega_{2} t) <\psi|(A_{1}(0) - A_{1}^{\dag}(0))|\psi>].
\label{momexp}
\end{eqnarray}
Inversion of the expressions in Eqs.(\ref{posexp}) and (\ref{momexp}) for the
measurement of
$(A_{1}(0)+A_{1}^{\dag}(0))$ and $(A_{1}(0)-A_{1}^{\dag}(0))$ on commuting operators
$q_{1}(t)$ and $p_{2}(t)$ is given by
\begin{eqnarray}
\left[ \begin{array}{c}
     <A_{1}(0)+A_{1}^{\dag}(0)> \\ <A_{1}(0)-A_{1}^{\dag}(0)> \end{array} \right]
& = &
\frac{-1}{\cos ((\omega_{1}-\omega_{2})t)}
\left| \begin{array}{cc}
       -\cos \omega_{2} t & i\sin \omega_{1} t \\
       -i\sin \omega_{2} t & \cos \omega_{1} t
       \end{array} \right| \times \nonumber \\
& & \left[ \begin{array}{c}
       (\sqrt{2m_{1}\omega_{1}}/c)<q_{1}(t)> \\
       (i \sqrt{2}/(\sqrt{m_{2} \omega_{2}} s))<p_{2}(t)> \end{array} \right].
\label{inversion}
\end{eqnarray}
The rotation $\vec{A}=R(\theta)\cdot\vec{a}$ corresponds to the system/meter entanglement
that defines a Naimark extension \cite{helstrom}. In this construction, a meter Hilbert
space $\cal{H}_{M}$ is combined with the system space $\cal{H}_{S}$ to define the space
$\cal{H}_{M} \otimes \cal{H}_{S}$ in which relevant operators commute. For the case of
position and momentum, commuting operators can be defined as $Q_{T}=Q_{1}-Q_{2}$ and
$P_{T}=P_{1}+P_{2}$, where the subscripts $1$ and $2$ correspond to system and meter,
respectively. Information of the system position and momentum is contained in the
orthonormal eigenstates $\{|\xi, \eta>\}$ of $(Q_{T},P_{T})$ in $\cal{H}_{M} \otimes
\cal{H}_{S}$. Mathematically, the projection property of a Naimark extension is defined
in terms of density operators $\rho_{S+M} = |\xi,\eta><\xi,\eta|$ and $\rho_{M}=|0><0|$,
which correspond to Naimark and meter states, respectively. Projection to a system
coherent state \cite{klauder} is then given by
\begin{equation}
|\alpha><\alpha| = Trace_{M}(\rho_{S+M}\rho_{M}),
\label{harmtrace}
\end{equation}
where $\alpha = \sqrt{m \omega /2}(q+ip/(m\omega))$ for a system of mass $m$, position
$q$, and momentum $p$; and where $Trace_{M}$ is the trace over meter states. The
projection property in Eq.(\ref{harmtrace}), a defining characteristic of Naimark
extensions, is the basis for entangled simultaneous measurement of position and momentum.
In a harmonic oscillator model of the measurement, Arthurs and Kelly
\cite{arthurskelly} observed that the entanglement of the system with two vacuum meters
resulted in the collapse of the system to a coherent state upon position and momentum
measurement on meters. However, the single system/meter entanglement above,
$\vec{A}=R(\theta) \cdot \vec{a}$, is sufficient to simultaneously measure position
and momentum. Of course in this case the system is projected onto an eigenstate of the
measured operator rather than a coherent state. This realization of the Naimark
extension, which is the minimum possible for entangled simultaneous measurement, is
used in this paper.

In Appendix~A the above construction is extended to a second quantized system described
by an angular momentum basis set $\{|jm>; j=0,1,\ldots;m=-j,\ldots,j\}$. It is shown
that, by entangling the system to two ancillary meters in vacuum states $|0>$, where the
vacuum is defined with zero occupation of the states $\{|jm>\}$, the original system angular
momentum components are simultaneously measured. Extended operators $\tilde{J}_{x}^{(1)}$,
$\tilde{J}_{y}^{(2)}$, and $\tilde{J}_{z}^{(3)}$, corresponding to the original system $(1)$
and meters $(2)$ and $(3)$, have a role similar to $Q_{T}$ and $P_{T}$ defined above.
These operators commute and project onto the system operators $J_{x}$, $J_{y}$, and
$J_{z}$ upon meter measurement. An analogy to the Naimark projection property in
Eq.~(\ref{harmtrace}) is also described.

The examples in this section and Appendix~A demonstrate the difference between the
formulation of entangled simultaneous measurement in first and second quantized
systems. In the former case, which has found applications in quantum optics
\cite{yuen}-\cite{yariv}, the entangling hamiltonian in Eqs.~(\ref{ham0}) and
(\ref{hamint}) is dependent on the measured position and momentum operators.
In a second quantized system the entangling interaction is independent of the
measured operators, which are represented as bilinear functions of creation and
annihilation operators (see Eqs.(\ref{a8})-(\ref{a10})). Because of this
property, the structure of the Naimark extension depends only on the {\em algebra}
represented by the observables. The minimal extension requires a
sufficient number of independent meters for the measurement of non-commuting
operator sets. For example, the Naimark extension of the combined galilean algebra
of position $Q_{i}$, momentum $P_{i}$, and angular momentum $J_{i}$, $i=x,y,z$,
appropriate for a non-relativistic particle \cite{weinberg}, is derived from the
commutation relations $j,k,m=x,y,z$,
\begin{equation}
\begin{array}{lll}
\left[Q_{k},P_{j} \right] = i \delta_{kj}  &
\left[J_{k},Q_{j} \right] = i \epsilon_{kjm} Q_{m} &
\left[J_{k},P_{j} \right] =i  \epsilon_{kjm} P_{m}  \nonumber \\
\left[J_{k},J_{j} \right] =i  \epsilon_{kjm} J_{m} &
\left[Q_{k},Q_{j} \right]= 0  &
\left[P_{k},P_{j} \right]= 0.
\end{array}
\label{galilean}
\end{equation}
A second quantized Naimark extension for the above algebra consists of three
distinguishable particles upon which mutually commuting operator pairs
$(J_{i},P_{i})$,$i=x,y,z$, are measured. All three commuting components
of the position vector $Q_{i}$,$i=x,y,z$, are measured on a fourth particle.
The four entangled particles of the minimum Naimark extension result from the
mutually commuting subsets of the algebra in Eq.(\ref{galilean}). This
construction is generalized to the Poincare algebra in the next section.
\section{Relativistic Systems}
The previous section contains constructions for entangled simultaneous measurement
of position, momentum, and angular momentum with quantum meters. The relativistic
generalization of these results is the entangled measurement of operators in
the Poincare algebra, from which particle representations satisfying Lorentz
invariance are obtained \cite{weinberg}. It is suggested that the Naimark extension
of this algebra is the basis for a larger structure than the usual particle
spin representations, and is the source of the quark/lepton multiplets.
Appendix~B contains the formulation of entangled simultaneity for fermions
described by second quantized Dirac fields. It is shown that $n$ non-commuting
operators $\{ \Theta_{k}, k=1,\ldots,n \}$ are simultaneously measurable
by the system entanglement with $(n-1)$ fermions. The Naimark-extended state
for this measurement is $|\phi>_{(1)}|0>_{(2)} \ldots |0>_{(n)}$, in which the ancillary
particles $(j), j=2,\ldots,n$ are in the vacuum state.

The relativistic generalization of position and momentum is the Poincare algebra of
operators involving the generators of Lorentz transformations. The nine operators,
not including the hamiltonian, can be grouped into three 3-vectors; momentum,
\begin{equation}
P_{j} = i \int d\vec{x} \psi^{\dag}(\vec{x},t) \partial_{j} \psi(\vec{x},j),
\label{relmom}
\end{equation}
angular momentum, $J_{k} = i \epsilon_{kjm}J^{jm}$, $k,j,m=1,2,3$, with
\begin{equation}
J^{jm} = i \int d\vec{x} \psi^{\dag}(\vec{x},t) (x^{j} \partial^{m} -
              x^{m} \partial^{j} - i {\cal J}^{jm} ) \psi(\vec{x},t),
\label{relang}
\end{equation}
and boost generators,
\begin{equation}
K_{j} = \int d\vec{x} \psi^{\dag}(\vec{x},t) (i t \partial_{j} + i x_{j} \partial_{0}
                      + {\cal K}_{j} ) \psi(\vec{x},t),
\label{relboost}
\end{equation}
with (assuming Pauli matrices $\sigma_{k}$, $k=1,2,3$)
\begin{equation}
{\cal J}^{jm} = i \epsilon^{jmk} \left[ \begin{array}{cc}
\sigma_{k} & 0 \\ 0 & \sigma_{k}
\end{array} \right],
\label{matrixj}
\end{equation}
and
\begin{equation}
{\cal K}_{j} = i/2 \left[ \begin{array}{cc}
\sigma_{j} & 0 \\ 0 & - \sigma_{j}
\end{array} \right].
\label{matrixk}
\end{equation}
The boost operators in Eq.(\ref{relboost}) contain the average position of particle
energy through the integral $\int d\vec{x} \vec{x} \psi^{\dag} H \psi $. The commutation
relation $[P_{k},K_{j}]= i \delta_{jk} H$ further suggests the identification of
$\vec{K}$ with position. However the boost operators are not hermitian. Consequently,
in order to describe the Poincare algebra with measurable observables, the real
and imaginary parts are separately defined as $K_{j} = K_{j}^{1} + i K_{j}^{2}$ with
$K_{j}^{i}$, $i=1,2$, hermitian. The operator $K_{j}^{1}$ is the physical location
of the particle energy, and $K_{j}^{2}$ reflects spin state effects in the boost
operator.

The operators in Eqs.(\ref{relmom})-(\ref{matrixj}) fit the model in Eq.(\ref{b6})
for which a second quantized Naimark extension is discussed in Appendix~B. The
complete Poincare algebra in terms of boost operators $K_{j}^{i}$, $i=1,2$, is
given by
\begin{eqnarray}
\left[J_{j},J_{k}\right]=i\epsilon_{jkm}J_{m}, &
\left[J_{j},K_{k}^{1}\right]=i\epsilon_{jkm}K_{m}^{1}, &
\left[K_{i}^{2},P_{j}\right]=0, \nonumber \\
\left[J_{j},P_{k}\right]=i\epsilon_{jkm}P_{m}, &
\left[J_{j},K_{k}^{2}\right]=i\epsilon_{jkm}K_{m}^{2},&
\left[K_{i}^{2},H\right]=0,  \nonumber \\
\left[K_{j}^{1},H\right]=iP_{j},&
\left[J_{i},H\right]=0, &
\left[P_{i},H\right]=0,  \nonumber \\
\left[K_{j}^{1},P_{k}\right]=i\delta_{jk} H, &
\left[K_{j}^{1},K_{k}^{2}\right] + \left[K_{j}^{2},K_{k}^{1}\right]= 0, & \nonumber \\
& \left[K_{j}^{1},K_{k}^{1}\right] - \left[K_{j}^{2},K_{k}^{2}\right]=
-i \epsilon_{jkm}J_{m}.  &
\label{poincare}
\end{eqnarray}
A minimal embedding of the Poincare algebra for massive fermions uses the commutations
$[J_{j},K_{j}^{1}]=[P_{j},K_{j}^{2}]=0$,$j=1,2,3$, in Eq.~(\ref{poincare}). Three
independent fields $\psi_{j}$ are sufficient for $(J_{j},K_{j}^{1})$, $j=1,2,3$,
measurements; and on an additional three fields $\psi_{j}^{\prime}$ the operators
$(P_{j},K_{j}^{2})$, $j=1,2,3$, are measured. The hamiltonian operator,
satisfying $[H,P_{j}]=[H,K_{j}^{2}]=0$, could be measured on any of the
$\psi_{j}^{\prime}$ fields. Because of the time derivative in the boost operator
in Eq.~(\ref{relboost}), the energy density position corresponds to the original
mass (before entanglement, $m_{j}$ in Eq.(\ref{b1})) rather than the
system mass ($M_{1}$ in Eq.(\ref{b2})). This suggests that the three fields
$\psi_{j}$ used to measure boost operators should have the same mass as the
original system, which implies constraints on the entanglement in
Eqs.~(\ref{b1})-(\ref{b3}).

The Naimark embedding of the Poincare algebra suggests an underlying explanation
for the generation structure of quarks and leptons. The fermions $(u,c,t)$,
$(d,s,b)$, $(e,\mu,\tau)$, and $(\nu_{e},\nu_{\mu},\nu_{\tau})$ are triplets
within which particles differ only by mass. A realization of the Naimark extension
is obtained by identifying massive quarks and leptons with the fields
$\psi_{j} (\psi_{j}^{\prime})$, $j=1,2,3$, to obtain the mapping
\begin{equation}
\left[ \begin{array}{ccc}
f^{1}_{L} & f^{2}_{L} & f^{3}_{L} \\
f^{1}_{R} & f^{2}_{R} & f^{3}_{R}
\end{array} \right] \longrightarrow
\left[ \begin{array}{ccc}
(J_{1},K_{1}^{1}) & (J_{2},K_{2}^{1}) & (J_{3},K_{3}^{1}) \\
(P_{1},K_{1}^{2}) & (P_{2},K_{2}^{2}) & (P_{3},K_{3}^{2})
\end{array} \right]
\label{fermions}
\end{equation}
where $f^{i} = (u,c,t),(d,s,b)$, and $(e,\mu,\tau)$. The entanglement between
different generations of left and right-handed fermions is the result
of spontaneous symmetry breaking of the vacuum by Higgs particles. As described
in Ref.~\cite{peskin}, the entanglement for quarks results from the
diagonalization of the coupling to Higgs particles expressed as rotations
$\vec{f}^{\prime}_{R} = W_{u(d)} \cdot \vec{f}_{R}$ and
$\vec{f}^{\prime}_{L} = U_{u(d)} \cdot \vec{f}_{L}$, where $u(d)$ corresponds to
{\em up (down)} quarks, and $\vec{f} = (f^{1},f^{2},f^{3})^{T}$ corresponds to
the triplets $(u,c,t)$ and $(d,s,b)$. The Cabibbo-Kobayashi-Maskawa matrix,
$V=U_{u}^{\dag}U_{d}$, is the only observable (other than mass) arising from
fermionic mixing in the standard model \cite{kobayashimaskawa}. Mass generation
from a non-zero Higgs vacuum expectation provides the
entanglement connecting left- and right-handed particles that completes
the fermionic Naimark extension.

Massless fermions form representations of a reduced Poincare algebra
\cite{weinberg}. For left-handed particles (like massless neutrinos) with
fixed $+1/2$ helicity, the only measurements
required in an inertial frame are boosts $K_{j}^{1}$ and momentum $P_{j}$
operators. From the commutation $[P_{j},K_{i}^{1}]=0$ for $i \neq j$,
the operator correspondence of left-handed fermions
\begin{equation}
(\nu_{e},\nu_{\mu},\nu_{\tau}) \longrightarrow
((P_{1},K_{2}^{1}),(P_{2},K_{3}^{1}),(P_{3},K_{1}^{1})),
\label{neutrinos}
\end{equation}
is a sufficient Naimark extension.
\section{Conclusions}
In this paper it is suggested that quantum mechanics is incomplete, and the
complete formulation is relevant at relativistic energies. A complete
quantum system is in an entangled state with the property that the entire
observable phase space of non-commuting operators is measurable.

The definition of a relativistic fermion as a minimally entangled system
for the representation of the Poincare algebra is examined. It is shown
that the generation structure of leptons and quarks fits this definition
with entanglement provided by Higgs couplings as observed in the
Cabibbo-Kobayashi-Maskawa matrix. In the high energy limit, the
primary structure is not a particle, but rather a system of six
entangled particles upon which the complete phase space is represented.
This is a generalization of the usual group theoretical particle
representation arising from Lorentz invariance of the lagrangian.
\appendix
\section{Entanglement for Angular Momentum}

In this appendix we construct the Naimark extension for angular
momentum measurement in a second quantized system described by
angular momentum states $\{|jm>;j=0,\ldots;m=-l,\ldots,l\}$ with a
non-interacting hamiltonian given by
\begin{equation}
H=\sum_{j=0}^{\infty}\sum_{m=-j}^{j}\sum_{k=1}^{3}
E_{k}(j) a_{jm}^{(k)\dag}a_{jm}^{(k)},
\label{a1}
\end{equation}
where $a_{jm}^{(k)}$, $k=1,2,3$, are the annihilation operators for three
independent systems satisfying commutation relations
\begin{eqnarray}
\left[a_{jm}^{(k)},a_{j^{\prime}m^{\prime}}^{(k^{\prime}) \dag}\right] =
\delta_{j j^{\prime}} \delta_{m m^{\prime}} \delta_{k k^{\prime}}, &
\left[a_{jm}^{(k)},a_{j^{\prime}m^{\prime}}^{(k^{\prime})}\right] = 0.
\label{a2}
\end{eqnarray}
The system/meter operators are defined by a rotation of the original system
as
\begin{equation} \left[
\begin{array}{c} A_{jm}^{(1)} \\ A_{jm}^{(2)} \\ A_{jm}^{(3)} \end{array}
\right] = R \cdot  \left[
\begin{array}{c} a_{jm}^{(1)} \\ a_{jm}^{(2)} \\ a_{jm}^{(3)} \end{array}
\right],
\label{a3}
\end{equation}
where $R$ is a $|jm>$-independent rotation matrix. Substitution of Eq.~(\ref{a3})
into Eq.~(\ref{a1}), with $E=diag(E_{1}(j),E_{2}(j),E_{3}(j))$, yields a
system/meter hamiltonian given by
\begin{equation}
H=\sum_{j=0}^{\infty}\sum_{m=-j}^{j}\sum_{k=1}^{3}\sum_{k^{\prime}=1}^{3}
A_{jm}^{(k) \dag} D_{kk^{\prime}}(j) A_{jm}^{(k^{\prime})},
\label{a4}
\end{equation}
where
\begin{equation}
D(j) = R E(j) R^{\dag}.
\label{a5}
\end{equation}
The expression in Eq.~(\ref{a4}) can be written as a non-interacting hamiltonian
(terms with $k=k^{\prime}$),
\begin{equation}
H_{0}=\sum_{j=0}^{\infty}\sum_{m=-j}^{j}\sum_{k=1}^{3}
A_{jm}^{(k) \dag} D_{kk}(j) A_{jm}^{(k)},
\label{a6}
\end{equation}
and an interaction term
\begin{eqnarray}
H_{int} & = & \sum_{j=0}^{\infty}\sum_{m=-j}^{j}
 [D_{12}(j) A_{jm}^{(1) \dag} A_{jm}^{(2)} +
 D_{13}(j) A_{jm}^{(1) \dag} A_{jm}^{(3)} \nonumber \\  & + &
 D_{23}(j) A_{jm}^{(2) \dag} A_{jm}^{(3)}]   +  h.c.
\label{a7}
\end{eqnarray}
Assume that the operators $A_{jm}^{(1)}$ and $A_{jm}^{(k)}$, $k=2,3$, correspond
to the system and meters, respectively, and consider the commuting angular
momentum operators for the {\em original} independent systems in Eq.(\ref{a1}),
\begin{equation}
\tilde{J}_{x}^{(1)}=\sum_{j}\sum_{mm^{\prime}} a_{jm}^{(1) \dag}(t)
                    a_{jm^{\prime}}^{(1)} (t) <jm|{\cal J}_{x} |jm^{\prime}>,
\label{a8}
\end{equation}
\begin{equation}
\tilde{J}_{y}^{(2)}=\sum_{j}\sum_{mm^{\prime}} a_{jm}^{(2) \dag}(t)
                    a_{jm^{\prime}}^{(2)} (t) <jm|{\cal J}_{y} |jm^{\prime}>,
\label{a9}
\end{equation}
and
\begin{equation}
\tilde{J}_{z}^{(3)}=\sum_{j}\sum_{mm^{\prime}} a_{jm}^{(3) \dag}(t)
                    a_{jm^{\prime}}^{(3)} (t) <jm|{\cal J}_{z} |jm^{\prime}>,
\label{a10}
\end{equation}
where ${\cal J}_{\alpha}$, $\alpha = x,y,z$, are matrix representations of the
angular momentum components. The {\em tilde} notation in Eqs.~(\ref{a8})-(\ref{a10})
is used to emphasize the distinction between the operators in the entangled
space and the system. The time dependence of $a_{jm}^{(k)}$, given by
$a_{jm}^{(k)} (t) = e^{-i E_{k} (j) t} a_{jm}^{(k)} (0)$, indicates that
the operators in Eqs.~(\ref{a8})-(\ref{a10}) are time independent. The substitution
of Eq.~(\ref{a3}) into Eqs.~(\ref{a8})-(\ref{a10}), and evaluation in the
state $|\psi>_{(1)}|0>_{(2)}|0>_{(3)}$ in which the meters $(2)$ and $(3)$ are
in vacuum states and the system $(1)$ is in the state $|\psi>$, results in the
expectation values,
\begin{eqnarray}
<\tilde{J}_{x}^{(1)}> & = & \sum_{j} \sum_{mm^{\prime}}
<\psi|A_{jm}^{(1) \dag}(t) A_{jm^{\prime}}^{(1)}(t)|\psi>
<jm|{\cal J}_{x}|jm^{\prime}> |R_{11}|^{2}  \nonumber \\ & = &
|R_{11}|^{2} <\psi|J_{x}^{(1)}|\psi>,
\label{a11}
\end{eqnarray}
\begin{eqnarray}
<\tilde{J}_{y}^{(2)}> & = & \sum_{j} \sum_{mm^{\prime}}
<\psi|A_{jm}^{(1) \dag}(t) A_{jm^{\prime}}^{(1)}(t)|\psi>
<jm|{\cal J}_{y}|jm^{\prime}> |R_{12}|^{2} \nonumber \\ & = &
|R_{12}|^{2} <\psi|J_{y}^{(1)}|\psi>,
\label{a12}
\end{eqnarray}
and
\begin{eqnarray}
<\tilde{J}_{z}^{(3)}> & = & \sum_{j} \sum_{mm^{\prime}}
<\psi|A_{jm}^{(1) \dag}(t) A_{jm^{\prime}}^{(1)}(t)|\psi>
<jm|{\cal J}_{z}|jm^{\prime}> |R_{13}|^{2} \nonumber \\ & = &
|R_{13}|^{2} <\psi|J_{z}^{(1)}|\psi>.
\label{a13}
\end{eqnarray}
The expressions in Eqs.~(\ref{a11})-(\ref{a13}) demonstrate the simultaneous
measurement of system angular momentum components on the commuting operators
$\tilde{J}_{x}^{(1)}$, $\tilde{J}_{y}^{(2)}$, and $\tilde{J}_{z}^{(3)}$. The
Naimark projection property corresponding to Eq.~(\ref{harmtrace}) is given by
\begin{equation}
J_{i}^{(1)} = Trace_{(2)(3)} \left[ \frac{\tilde{J}_{i}^{(i)} \rho_{M}}
                                         {|R_{1i}|^{2}} \right],
                                         i=1,2,3,
\label{a14}
\end{equation}
where $J_{i}^{(i)}$ is the $i^{th}$ component of angular momentum for the system,
$\rho_{M} = |0><0|$ is the density matrix for the product vacuum state
$|0> = |0>_{(2)} |0>_{(3)}$, and the trace is over the Hilbert space for meters
$(2)$ and $(3)$.
\addtocounter{section}{+0}
\section{Entanglement for Dirac Fields}

Consider the Dirac hamiltonian for $n$ independent fermionic fields,
\begin{equation}
H=\sum_{j=1}^{n} \int d\vec{x} (\pi_{j} \gamma^{0} \vec{\gamma} \cdot
                  \vec{\nabla} \psi_{j} + m_{j} \pi_{j} \gamma^{0} \psi_{j}),
\label{b1}
\end{equation}
where $\psi_{j}(\vec{x},t)$, $j=1,\ldots,n$ are second quantized Dirac spinors,
$\gamma^{\mu}$ are Dirac matrices (Ref.~\cite{weinberg} notation), and
$\pi_{j}=i \psi_{j}^{\dag}$ is the conjugate momentum to the field $\psi_{j}$.
The rotation of the vector of fermionic fields, $\vec{\psi}^{\dag} =
(\psi_{1}^{\dag},\ldots,\psi_{n}^{\dag})$, given by $\vec{\Psi} = R \cdot
\vec{\psi}$, defines the system/meter Hilbert spaces. Substitution into Eq.~(\ref{b1})
results in a hamiltonian $H=H_{0}+H_{int}$ with
\begin{equation}
H_{0}=\sum_{j=1}^{n} \int d\vec{x} (\Pi_{j} \gamma^{0} \vec{\gamma} \cdot \vec{\nabla}
\Psi_{j} + M_{j} \Pi_{j} \gamma^{0} \Psi_{j}),
\label{b2}
\end{equation}
and
\begin{equation}
H_{int}= \int d\vec{x} (\vec{\Pi} \gamma^{0}
({\cal M} - diag(M_{1},\ldots,M_{n}))\vec{\Psi}),
\label{b3}
\end{equation}
with $\Pi_{j} = i \Psi_{j}^{\dag}$, $m=diag(m_{1},\ldots,m_{n})$,
${\cal M}=RmR^{\dag}$, and $M_{j} = {\cal M}_{jj}$.

The free fields $\psi_{j}$, $j=1,\ldots,n$, satisfy the equation,
\begin{equation}
\psi_{j}(\vec{x},t)=e^{iH_{0}(j)t} \psi_{j}(\vec{x},0) e^{-iH_{0}(j)t},
\label{b4}
\end{equation}
where $H_{0}(j)$ is the $j^{th}$ term in Eq.~(\ref{b1}). The second quantized field in
Eq.~(\ref{b4}) is written in terms of the vector $\vec{\Psi}(\vec{x},0)$, at time
zero, as
\begin{equation}
\psi_{j} (\vec{x},t) = e^{iH_{0}(j) t} (R^{\dag} \vec{\Psi}(\vec{x},0))_{j}
e^{-iH_{0}(j) t}.
\label{b5}
\end{equation}
Consider the entangled simultaneous measurement of field operators,
\begin{equation}
\Theta_{k} = \int d\vec{x} \psi_{1}^{\dag} (\vec{x},t) \theta_{k} (\vec{x})
                           \psi_{1}(\vec{x},t), k=1,\ldots,n,
\label{b6}
\end{equation}
where $\theta_{k}(\vec{x})$ are $4 \times 4$ $\vec{x}$-dependent operators. The
substitution of Eq.~(\ref{b5}) into Eq.~(\ref{b6}), with $\psi_{1}$ replaced with
$\psi_{k}$, results in the expression,
\begin{equation}
\tilde{\Theta}_{k}^{(k)} (t) = e^{iH_{0}(k) t}
\int d\vec{x} (\vec{\Psi}^{\dag}(\vec{x},0)R^{\dag})_{k} \theta_{k}(\vec{x})
(R\vec{\Psi}(\vec{x},0))_{k} e^{-i H_{0} (k) t}.
\label{b7}
\end{equation}
Note that the operators $\tilde{\Theta}_{k}^{(k)}$,$k=1,\ldots,n$, are mutually
commuting from the independence of the fields $\psi_{k}$. The commuting set of
operators, $\tilde{\Theta}_{k}^{(k)} (0) = e^{-i H_{0} (k) t}
\tilde{\Theta}_{k}^{(k)} (t) e^{i H_{0} (k) t}$, evaluated in the state
$|\phi>_{(1)}|0>_{(2)} \ldots |0>_{(n)}$, where $(k)$ corresponds to the system
with field $\Psi_{k}$, results in the expression,
\begin{equation}
<\tilde{\Theta}_{k}^{(k)} (0)> = <\phi|\int \Psi_{1}^{\dagger} (\vec{x},0)
\theta_{k}(\vec{x}) \Psi_{1}(\vec{x},0) |\phi> |R_{1k}|^2.
\label{b8}
\end{equation}
The expression in Eq.~(\ref{b8}) represents the entangled simultaneous measurement
of $\Theta_{k}$, $k=1,\ldots,n$, in the state $|\phi>_{(1)}$ through
the Naimark extension to $n$ fermions. The equation is a relativistic generalization
to an arbitrary set of operators of Eqs.~(\ref{a11})-(\ref{a13}) for non-relativistic
angular momentum. The Naimark projection property in Eqs.~(\ref{harmtrace}) and
(\ref{a14}) is generalized to the relativistic case by the condition
\begin{equation}
\Theta_{k} = Trace_{(2)(3)\ldots (n)} \left[
\frac{\tilde{\Theta}_{k}^{(k)} (0) \rho_{M}}{|R_{1k}|^{2}} \right],
k=1,\ldots,n,
\label{b9}
\end{equation}
where $\rho_{M} = |0><0|$ for the meter state $|0>=|0>_{(2)} \ldots |0>_{(n)}$.
\addtocounter{section}{+0}
\section{Quantum Logical Implications}
Measurement of non-commuting operators in a quantum system is the cause of
the non-distributivity in the von Neumann subspace lattice. One approach to the
inconsistencies and paradoxes arising from this property is to restrict the
interpretation of these measurements as simultaneous only if
joint measurements occur on an entangled system at a fixed space-time
location. In this appendix, this procedure for the measurement of incompatible
observables is included in the subspace lattice of logical propositions. It
is shown that the Naimark-extended von Neumann lattice is distributive for a
simple example of spin-1/2 component measurement.
\subsection{Entangled Simultaneous Spin Measurement}
In this section the entangled simultaneous measurement of spin-1/2 operators
$S_{x}$ and $S_{z}$ is demonstrated by the coupling to an independent
spin-1/2 meter. This derivation is a simplified version of the mechanism proposed
by Levine and Tucci \cite{levinetuccib}.

Consider two independent spin-1/2 systems $S$ and $R$ defined by commuting spin-1/2
operators $S_{j}$ and $R_{j}$, $j=1,2,3$, which are both expressed as Pauli
matrices $\sigma_{j} /2$, $(\hbar = 1)$. Assume an entangling hamiltonian given by
\begin{equation}
H=k S_{x} R_{z},
\label{c11}
\end{equation}
where $k$ is an arbitrary coupling constant. The evolution operator $U=e^{-iHt}$,
corresponding to the hamiltonian $H$ in Eq.~(\ref{c11}), is given by
\begin{equation}
U = (c - 4is S_{x} R_{z})
\label{c12}
\end{equation}
where $c=\cos(kt/4)$ and $s=\sin(kt/4)$. The evolution of operators $S_{z}(t)$
and $R_{x}(t)$ in the Heisenberg picture determines the entangled spin component
measurement at time $t$ on the $S$ and $R$ systems. The evolved, entangled,
and commuting operators are given from Eq.~(\ref{c12}) by
\begin{equation}
R_{x} (t) = [ (c^{2}-s^{2})R_{x} + 4csS_{x}R_{y}],
\label{c13}
\end{equation}
and
\begin{equation}
S_{z} (t) = [ (c^{2}-s^{2})S_{z} - 4csS_{y}R_{z}].
\label{c14}
\end{equation}
Assume that the $R$-system is aligned along the $+y$ direction in the state
$|r>=|+1/2>_{y}^{R}$, and note that the expectations $<r|R_{z}|r>$ and
$<r|R_{x}|r>$ vanish, to obtain the projection of Eqs.~(\ref{c13}) and
(\ref{c14}) to $S$ system components,
\begin{equation}
<r|R_{x}(t)|r>=2sc S_{x},
\label{c15}
\end{equation}
and
\begin{equation}
<r|S_{z}(t)|r>=(c^{2}-s^{2}) S_{z}.
\label{c16}
\end{equation}
The operators $R_{x} (t)$ and $S_{z} (t)$ provide an entangled simultaneous
measurement of $S_{x}$ and $S_{z}$. The above projections are typically expressed
as a partial trace over the meter Hilbert space of the product
of the meter density operator $\rho_{M} = |r><r|$ and the relevant
extended-space operator \cite{helstrom}. For
example, Eqs.~(\ref{c15}) and (\ref{c16}) are expressed as
\begin{equation}
S_{x} = \frac{1}{2sc} Trace_{R} [R_{x}(t) \rho_{M}],
\label{c17}
\end{equation}
and
\begin{equation}
S_{z}=\frac{1}{(c^{2}-s^{2})} Trace_{R} [S_{z} (t) \rho_{M}].
\label{c18}
\end{equation}
Mathematically, the $(S,R)$-Hilbert space ${\cal H}_{S} \otimes {\cal H}_{R}$
is the Naimark extension of the $S$-Hilbert space ${\cal H}_{S}$
\cite{naimarka}-\cite{helstrom}. Among the proposals in this paper is that the
Naimark extension is a logically consistent mechanism for joint quantum
measurements of non-commuting operators.
\subsection{Extended Subspace Lattice}
The essential property of an extended quantum system is seen in the lattice of
spin-1/2 measurements, which is defined in Refs.~\cite{hughesa} and \cite{hughesb}.
Consider
a spin-1/2 system $S$ and an apparatus that measures spin in the $x$ or $z$
directions with the result up $u$ or down $d$. As discussed in the previous
section, a spin-1/2 meter $R$ is entangled with $S$ such that the $R$-vacuum
expectation value directly reveals the corresponding spin operator of $S$.
Denote by $\alpha \beta \gamma$, with $\alpha \in \{r,s\}$, $\beta \in \{x,z\}$,
and $\gamma \in \{u,d\}$, the projected subspace for the measurement on the
system $\alpha$ of spin value $\gamma$ along direction $\beta$. For example,
$sxu$ defines the subspace generated by the $S_{x}$ eigenvector with eigenvalue
$u$ in system $S$. A subspace lattice of these outcomes on the system $S$ is
shown in Fig.~1, where $\vee (\wedge)$ represents {\em or} ({\em and})
connectives on the measurement outcomes \cite{hughesa}. The bold lines in Fig.~1
correspond to measurement outcomes, and the connecting dotted lines denote
inclusion into a higher dimensional subspace. The horizontal/vertical and
$45^{\circ}$-rotated coordinate systems correspond to $S_{x}$ and $S_{z}$
measurements (either $u$ or $d$), respectively.

A defining feature of quantum lattices is a failure of the distributive property
of meet $(m)$ over join $(j)$, which is seen in Fig.~1 by the different outcomes
in
\begin{equation}
sxu \, m \, (szu \, j  \, szd) = sxu,
\label{c21}
\end{equation}
and
\begin{equation}
(sxu \, m \, szu)\, j \, (sxu \, m \, szd) = \{s\} \, j \, \{s\} = \{s\},
\label{c22}
\end{equation}
where $m$ and $j$ denote {\em meet} and {\em join} operations on the lattice, and
the notation
$\{\alpha\} = (\alpha x u \wedge \alpha x d) = (\alpha z u \wedge \alpha z d)$
and
$[\alpha] = (\alpha x u \vee \alpha x d) = (\alpha z u \vee \alpha z d)$
is used. The failure of the distributive property is due to the fact that the
observables $S_{x}$ and $S_{z}$ are incompatible; suggesting that the distributive
property will hold in an extended lattice with measurements on commuting operators.

The extended Naimark subspace lattice, which combines the $S$ and $R$ Hilbert
spaces, is shown in Fig.~2. The vertically-placed coordinates in each row
correspond to $S$ (top) and $R$ (bottom) Hilbert spaces that have been entangled
to provide simultaneous measurement of system $S$ components. A  mechanism for
the entangled simultaneous measurement, with $R$ in a special initial state as
a meter, was discussed in Section~C.1. Consider a section of the extended lattice
referring to the simultaneous measurement of the $S$ system spin along the
$x$ and $z$ directions. The relevant lattice operations
\begin{eqnarray}
(sxu \wedge [r]) \, m \, [(szu \wedge [r])\, j \, (rzd \wedge [s])] &  = &
(sxu \wedge [r]) \, m \, ([r] \wedge [s]) \nonumber \\ &  = &  sxu \wedge [r],
\label{c23}
\end{eqnarray}
and
\begin{equation}
\begin{array}{lll}
[(sxu \wedge [r]) \, m \, (szu \wedge [r])] \, j \,
[(sxu \wedge [r]) \, m \, (rzd \wedge [s])] &  = & \\
(\{s\} \wedge [r]) \, j
\, (sxu \wedge rzd)   =  sxu \wedge [r], & &
\end{array}
\label{c24}
\end{equation}
are distributive. All other complex statements mixing $S_{x}$ on $S$ and
$S_{z}$ on $R$ are distributive in the lattice. For example, another
distributive expression is given by
\begin{equation}
\begin{array}{lll}
(sxu \wedge [r]) \, m \, [(\{s\} \wedge rzu) \, j \, (\{s\} \wedge rzd)]  = & &
 \\
\left[(sxu \wedge [r]) \, m \, (\{s\} \wedge rzu)\right] \, j \,
\left[(sxu \wedge [r]) \, m \, (\{s\} \wedge rzd)\right] & = &
\\ (\{s\} \wedge [r]). & &
\end{array}
\label{c25}
\end{equation}

In this appendix a distributive quantum theory is defined using special
measurement-dependent statements in cases of incompatible observables. The
appropriate measurements involve meters in special-purpose initial states
that are entangled with the system. In the body of the paper it was suggested
that such measurements are fundamental in the definition of relativistic
particle observables. The special meter initial state for particles is the
vacuum, the Naimark extension is the repeated fermionic generations, and
entanglement is observed in the Cabibbo-Kobayashi-Maskawa matrix and spontaneous
breaking of the electro-weak gauge symmetry.
\begin{figure}
\vspace{.0in}
\caption{Quantum subspace lattice for system $S$ spin measurements. Bold lines
are the measurement outcome subspaces. Connecting lines indicate inclusion
of lattice subspaces. {\em Figure attached in file Fig1.gif}.}
\end{figure}
\begin{figure}
\vspace{.0in}
\caption{Naimark-extended quantum lattice for entangled simultaneous spin-1/2
measurement with two systems $S$ and $R$. Subspace for system $S$ outcome is
above meter $R$ subspace. Note that $\wedge d$ and $d\wedge$ represent the
replacement of $u$ by $d$ in the outcome on the coordinate systems immediately
to the left. {\em Figure attached in file Fig2.gif}.}
\end{figure}
\pagebreak
\end{document}